\documentstyle[prl,aps,epsfig,multicol]{revtex}
\begin{document}

\title{Evidence for nodal quasiparticles in electron-doped cuprates from
penetration depth measurements}
\author{R. Prozorov and R. W. Giannetta}
\address{Loomis Laboratory of Physics, University of Illinois at Urbana-
Champaign, 1110 West Green St., Urbana, Illinois 61801}
\author{P. Fournier and R. L. Greene}
\address{Center for Superconductivity Research, Department of Physics,
University of Maryland, College Park, Maryland 20742 }
\date{submitted to Phys. Rev. Lett. February 18, 2000. Revised: May 16, 2000}
\maketitle

\begin{abstract}
The in-plane magnetic penetration depth, $\lambda ( T )$, was measured down to
0.4 K in single crystals of electron-doped superconductors,
Pr$_{1.85}$Ce$_{0.15}$CuO$_{4-\delta}$ (PCCO) and
Nd$_{1.85}$Ce$_{0.15}$CuO$_{4-\delta}$ (NCCO). In PCCO, the superfluid density
varies as $T^2$ from 0.025 up to roughly 0.3 $T/T_c$ suggestive of a d-wave
state with impurities. In NCCO, $\lambda ( T )$ shows a pronounced upturn for
$T < 4$ K due to the paramagnetic contribution of Nd$^{3+}$ ions. Fits to an
s-wave order parameter over the standard BCS range ($T/T_c$ = 0.32) limit any
gap to less than $\Delta_{min} (0)/T_c = 0.57$ in NCCO.  For PCCO, the absence
of paramagnetism permits a lower temperature fit and yields an upper limit of
$\Delta_{min} (0)/T_c = 0.2$. \end{abstract}


\begin{multicols}{2}
\narrowtext

There is by now a consensus that the hole-doped high-$T_c$ cuprates exhibit
$d-$wave pairing symmetry
\cite{dwave,hirschfeld,vanharlingen,tsuei,broun,hardy,kamal,jacobs,panagopoulos,covington}.
For electron-doped cuprates \cite{tokura} the issue remains unresolved. While
most theories for the mechanism of high temperature superconductivity are
insensitive to the sign of the carriers, some predict that \textit{n} and
\textit{p} type materials will have different pairing symmetry, making its
determination an important challenge \cite{rokhsar,abrikosov}. Early microwave
measurements of the penetration depth in NCCO were interpreted within an
$s-$wave model \cite{wu,anlage,andreone}. However, Cooper pointed out that the
power law dependence for $\lambda ( T )$ indicative of a nodal order parameter
could be masked by a large paramagnetic contribution from $Nd^{+3}$ ions
\cite{cooper}. Newer microwave measurements by Kokales {\it et al.}, performed
on the same sample used in this paper, have revealed an upturn and power-law
temperature dependence and are consistent with our data \cite{anlaged}.
Measurements of $\lambda ( T )$ using single grain boundary junctions
\cite{alff} have favored a gapped state. Some tunneling measurements favor an
$s-$wave order parameter, albeit with significant departures from an isotropic
weak coupling BCS picture \cite{ekino}, while others report a zero-bias
conductance peak \cite{hayashi,fournier,mourachkine}. Half-integral flux
indicative of d-wave pairing \cite{tsuei} was recently reported in tricrystal
experiments with both NCCO and PCCO films\cite{tsueincco}.

In this letter we report measurements of $\lambda ( T )$ down to 0.4 K in
single crystals of both NCCO and PCCO. Lower temperatures and higher
resolution combine to permit a more precise determination of the temperature
dependence of $\lambda ( T )$ than any previously reported. In NCCO, a large
paramagnetic contribution is observed below 4 K. In non-magnetic PCCO, we find
an overall $T^2$ variation of the superfluid density up to $T/T_c \approx
0.3$, suggesting the presence of nodal quasiparticles in the presence of
strong impurity scattering.

Single crystals of R$_{1.85}$Ce$_{0.15}$CuO$_{4-\delta}$ (R=Nd or Pr) were
grown using directional solidification technique and annealed in argon to
achieve optimal superconducting properties \cite{growth}. Penetration depth
was measured using an 11 MHz tunnel-diode driven LC resonator
\cite{degrift,carrington,prozorov2}. Samples were mounted on a movable
sapphire stage with temperature controllable from 0.4 K to 100 K. The low
noise level, $\Delta f_{min}/f_{0}\approx 5 \times 10^{-10}$, results in a
sensitivity of $\Delta \lambda \leq 0.5$ \AA\ for our samples [$0.5 \times 0.5
\times 0.02$ mm]. The large anisotropy of these materials
($\lambda_c/\lambda_{ab} \approx$ 30-80 \cite{anlage}) forces one to apply the
rf field perpendicular to the conducting planes. Otherwise, the frequency shift
will be dominated by changes of the interplane penetration depth, for which
there exists no straightforward connection to the pairing symmetry. A
semi-analytical solution for the rf susceptibility of a platelet sample of
square base $2w\times 2w$ and thickness $2d$ in this orientation was analyzed
in detail in Ref. \cite{prozorov2}. At low temperatures the frequency shift,
$\Delta f \equiv f(T)-f(0)$, is related to the change in penetration depth,
$\Delta \lambda \equiv \lambda (T)-\lambda (0)$, via $\Delta f=-G \Delta
\lambda$, with the calibration constant $G=V_s f_0/ \left[ 2 V_0 (1-N) R
\right]$, where $N$ is the effective demagnetization factor, $V_{s}$ is the
sample volume, $V_{0}$ is the effective coil volume, $f_0$ is the resonance
frequency in the absence of a sample and $R \approx 0.2 w$ is an effective
dimension \cite{prozorov2}. Although this result is similar to the known
solution for an infinite slab in parallel field \cite{hardy}, the effective
dimension $R$ differs significantly from $R=w/2$ obtained for an infinite bar.
This difference is due to penetration of the magnetic field from the top and
bottom surfaces. The sample and apparatus dependent constant $\Delta f_{0}
\equiv \ V_{s}f_{0}/\left[ 2 V_{0} (1-N) \right] $ is measured by moving the
sample out of the coil in situ. The overall calibration was tested with
samples of Nb, YBCO and BSCCO and gave $d\lambda/dT$ within 10 \% \ of
reported values \cite{hardy,kamal,jacobs}. In order to determine the
normalized superfluid density, $\rho_s \equiv \left[ \lambda (0)/\lambda (T)
\right]^2$, it is necessary to know the absolute magnitude of the penetration
depth, $\lambda (0)$. Measured values of $\lambda (0)$ in NCCO vary from 1000
- 2600 \AA\ \cite{wu,anlage,gollnik,nugroho}. In PCCO the only reported value,
$\lambda (0) \approx 1000$ \AA, was estimated from the measurements of the
lower critical field $H_{c1}$ and is less reliable due to demagnetization and
possible vortex effects \cite{hoekstra}. We recently developed a new technique
for determining $\lambda (0)$ from the frequency shift obtained by warming a
sample coated with a thin layer of Al above $T_c(Al)$. A detailed description
of this procedure will be published elsewhere \cite{prozorov}. This technique
applied to PCCO gave $\lambda (0) = 2500 \pm 100 \AA$ \cite{prozorov} which
will be used as an upper estimate in this paper.

\begin{figure}
 \centerline{\psfig{figure=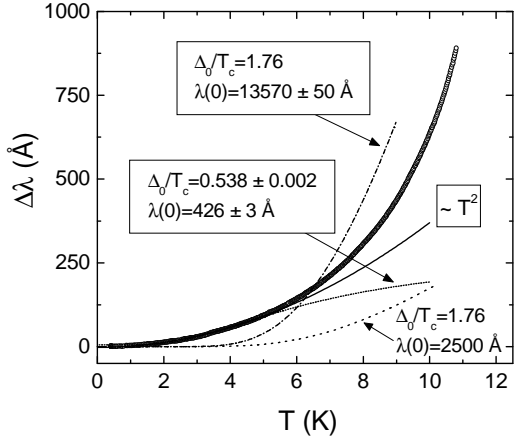,width=7cm,clip}}
 \caption{Low temperature variation of the penetration depth
 $\Delta \lambda ( T )$ in PCCO single crystal. Lines show fits to different
 models described in the text.}
 \label{p_lambda}
\end{figure}

In Fig.\ref{p_lambda} we plot the penetration depth for PCCO sample 1 along
with several fits. The fitting range was 5 K to assure validity of the low
temperature BCS expansion for an isotropic s-wave state, $\Delta \lambda =
\lambda (0) \sqrt{\pi \Delta (0)/2T}\exp{(-\Delta (0)/T)}$\cite{dwave}.  In
each case, the small negative offset $A= \lambda (0) - \lambda (0.4\ K)$ was
determined as a fit parameter. The solid line is a fit with 188 data points to
a power law, $\Delta \lambda$ = $A+B T^2$ with $B=3.70 \pm 0.01\ \AA/K^2$ and
$\chi^2 \approx 8.2$. The short dotted line shows the best fit to the BCS
s-wave expression. With both $\lambda (0)$ and $\Delta (0)$ as free parameters
we obtained $\Delta (0)/T_c = 0.538 \pm 0.002$ and $\lambda (0)=426 \pm 3$
\AA. The s-wave fit is somewhat worse than the power law ($\chi^2 \approx
17.6$) and gives an unrealistic value for $\lambda (0)$. The dash-dotted line
shows the s-wave fit where $\Delta (0)/T_c $ was fixed at the weak-coupling
BCS value (1.76). In this case an unrealistically large $\lambda (0)=13570 \pm
50$ \AA\ was obtained.  For comparison, the dotted line is a plot of the BCS
expression with $\Delta (0)/T_c = 1.76$ and set to a more realistic value of
$\lambda(0)=2500$ \AA.

If the order parameter is an anisotropic s-wave, then the minimum gap value
determines the low temperature asymptotic behavior. The BCS functional form
for $\Delta \lambda$ still holds, but with $\Delta_{min}$ replacing the
isotropic gap. The temperature range over which this asymptotic form is valid
is now reduced accordingly.  For an isotropic gap, $\Delta(0)/T_c=1.76$, the
range of validity in reduced temperature is $(T/T_c)_{max}$ =  $t_{max}
\approx 0.32$ \cite{dwave}. For an anisotropic gap, simple rescaling forces
the range of validity down to $t_{max} \approx 0.18 \Delta_{min}(0)/T_c$.
Without \textit{a priori} knowledge of $\Delta_{min}(0)/T_c$, we do not know
$t_{max}$ and so it is necessary to successively reduce the range until the
gap value obtained from the fit becomes independent of the range. Following
this procedure for PCCO, with $\lambda (0)=2500$ \AA\ fixed, we find that
$\Delta(min)/T_c$ extrapolates to $0.20 \pm 0.05$ as $t_{max} \to 0$. The same
procedure for Nb yields $\Delta(0)/T_c = 1.74 \pm 0.02$, as expected. This
stricter criterion would imply that any residual gap is less than $11 \%$ of
the isotropic BCS value.

The overall best fit for sample 1 was achieved for a $T^{2.25 \pm 0.01}$ power
law ($\chi^2 \approx 2.1$).  Although this may appear unphysical, it is
important to recall that the integer power laws expected for a nodal order
parameter strictly apply to the normalized superfluid density, $\rho_s$, and
not the measured quantity, $\Delta \lambda$.  If $\rho_s = 1 - c_n T^n$ then
$\Delta \lambda$ has corrections of order $T^{2n}$ and a fit to $\Delta
\lambda$ can result in an artificial intermediate power (e.g. 2.25).  This
distinction is clearly evident in high quality, untwinned YBCO above 10 K
\cite{carrington}.

\begin{figure}
 \centerline{\psfig{figure=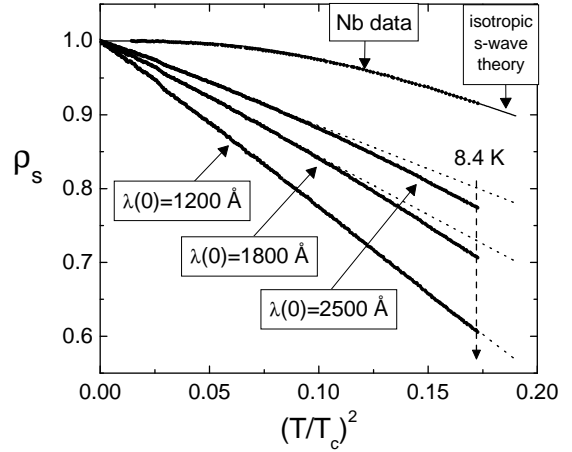,width=7.5cm,clip}}
 \caption{Superfluid density $\rho_s$ vs. $\left( T/T_c \right)^{2}$,
 calculated assuming $\lambda ( 0 )=$ 1000, 1550 and 2000 \AA. Dotted lines
indicate pure quadratic power law. Dash-dotted line shows isotropic $s-$wave
behavior.}
 \label{p_rho_lamb}
\end{figure}

In Fig. \ref{p_rho_lamb} we plot, for sample 1, normalized superfluid density,
$\rho_s \approx \left[ 1 + \Delta \lambda (T)/\lambda (0) \right]^{-2}$, vs. $(
T/T_c )^{2}$ for three choices of $\lambda (0)$ spanning the range of reported
values. $\lambda ( 0 )=$1200 \AA\
 yields $1-\rho_s \sim\ (T/T_c)^2$ up to 8.4 K, while larger choices for $\lambda ( 0 )$ reduce the range of pure quadratic behavior. For comparison we show data taken for Nb in the same apparatus. Up to to $T/T_c = 0.5$, the Nb data fits the low temperature BCS expansion perfectly with $\Delta (0)/Tc = 1.74 \pm 0.02$, giving us confidence that the measurement technique is sound.

\begin{figure}
 \centerline{\psfig{figure=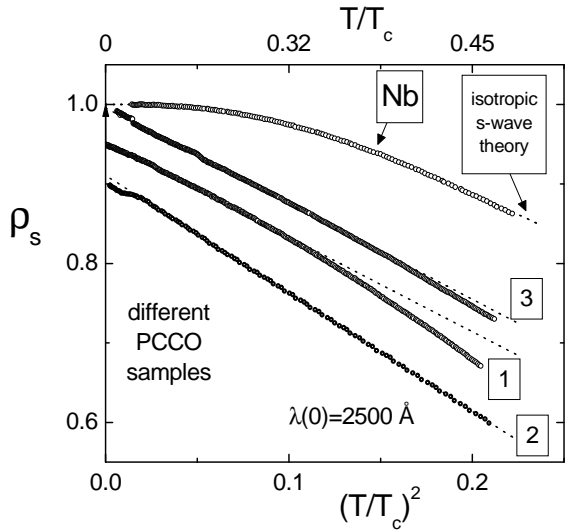,width=7.5cm,clip}}
 \caption{$\rho_s$ in three PCCO single crystals assuming $ \lambda (0)=2500$
 \AA\ (data offset for clarity). Dotted lines show $T^2$ fits at low temperatures. Also shown data
 for Nb and calculated isotropic s-wave curve.}
 \label{p_rho_samp}
\end{figure}

With an exponent of $n = 2.25$, sample 1 is our weakest candidate for a nodal
order parameter. Fig. \ref{p_rho_samp} shows data for samples 1, 2 and 3
(offset for clarity) and the same Nb data with a fit to the BCS form.  Samples
2 and 3 have power laws much closer to n = 2. In this plot we have chosen the
largest value of $\lambda (0)$ in order to cast the power law model in the most
unfavorable light (i.e., smallest range of pure quadratic behavior).  We
conclude that the superfluid density in PCCO is best described by a quadratic
power law variation with temperature.

There are currently two theories for a quadratic power law in d-wave
superconductors. Kosztin and Leggett showed that the divergence of the
effective coherence length near the nodes of a $d-$wave order parameter yields
$1-\rho_s \sim\ T^2$ due to nonlocal electrodynamics \cite{kosztin}.
Nonlocality is predicted to arise for the orientation used here (field
perpendicular to conducting planes) below $T_{nonlocal} \approx \xi (0) \Delta
(0)/\lambda (0)$, where $\xi(0) $ is the coherence length. In electron-doped
cuprates $T_{nonlocal} \approx$ 0.5 K to 2.5 K within our current knowledge of
superconducting parameters. Since we observe a quadratic temperature
dependence up to 8-10 K in some samples, nonlocality is unlikely to be the
source. A stringent test for nonlocality would require a comparison between
this data and $\lambda (T)$ obtained from the $H || ab$ plane orientation.
However, the $H || ab$ orientation involves the interplane penetration depth,
as discussed earlier.

\begin{figure}
 \centerline{\psfig{figure=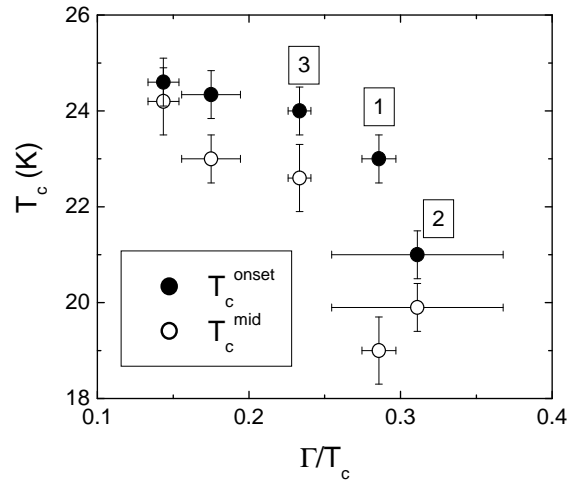,width=7.5cm,clip}}
 \caption{$T_c$ versus scattering parameter $\Gamma/T_c$ determined using 5 PCCO crystals.
 Filled symbols show $T_c^{onset}$ defined as the onset of diamagnetism. Open symbols
 show $T_c^{mid}$ defined as the inflection point on $\Delta \lambda (T)$.}
 \label{gamma}
\end{figure}

Impurity scattering in the unitary limit provides a more plausible explanation
for the quadratic dependence of $\rho_s ( T )$ \cite{hirschfeld,hardy}. In the
``dirty $d-$wave`` scenario, $\rho_s ( T )$ will cross over from a linear to
quadratic temperature dependence below $T^* \simeq 6 \ln{2} \gamma/\pi$, where
$\gamma \simeq 0.63 \sqrt{\Gamma \Delta (0)}$ and $\Gamma$ is a scattering
rate parameter, proportional to the impurity concentration \cite{hirschfeld}.
The slope $\left. {d\lambda /dT^2 } \right|_{T \to 0} \simeq \pi \lambda
(0)/\left(6 \gamma \Delta (0) \right)$. Casting this result in dimensionless
form gives: $\Gamma/T_c \simeq 0.28/\left[ -d \rho_s/d(T/T_c)^2 \right]$. In
Fig. \ref{gamma} we plot $T_c$ versus $\Gamma/T_c$ for all five samples
studied.  Samples 1,2 and 3 are marked. The transitions in some samples were
broad and two different criteria were used to estimate $T_c$ - onset of the
diamagnetic signal and the inflection point of the $\Delta \lambda (T)$ curve.
In general, the trend shows $T_c$ suppression with increased scattering rate
as expected for a d-wave state with impurity scattering \cite{hirschfeld}.
$\Gamma/T_c$ is at least 10 times larger than the scattering rate observed in
clean YBCO \cite{hardy,kamal}.

We now discuss measurements of $\lambda ( T)$ in NCCO. This compound has been
studied much more thoroughly than PCCO and was cited as the first evidence for
$s-$wave pairing in e-doped materials \cite{wu,anlage}. However, Nd$^{3+}$
ions introduce a large paramagnetic background and influence the measured
penetration depth \cite{cooper}. With magnetic permeability $\mu ( T
)=1+C/\left(\Theta+T \right)$, the measured penetration depth is given by
$\lambda ( T )=\lambda_L ( T ) \sqrt{\mu ( T )}$, where $\lambda_{L}$ is the
London penetration depth, $C$ is a Curie-Weiss constant, and $\Theta$ is the
characteristic temperature for antiferromagnetic interaction\cite{cooper}. To
fit the data,we take $\Theta=1.2$ K from neutron scattering \cite{boothroyd}
and specific heat \cite{maple} measurements. For $C$ we have chosen two
representative values $C=0.3$ and $0.05$, calculated assuming the effective
magnetic moment of Nd$^{3+}$ ions to be $2.4 \mu_B$ \cite{cooper} and $1
\mu_B$, respectively.

\begin{figure}
\centerline{\psfig{figure=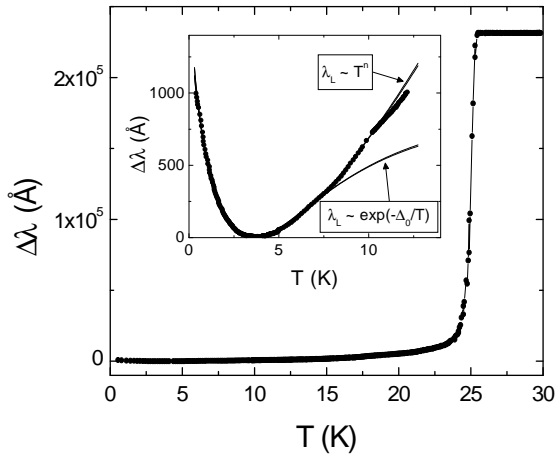,width=7.5cm,clip}}
 \caption{$\Delta \lambda ( T )$ in NCCO single crystal.
{\it Inset:} Low temperature range. Lines are the fits assuming a power law or
an $s-$wave temperature dependence for the London penetration depth $\lambda_L
(T)$ as described in the text.}
 \label{ncco}
\end{figure}

Figure \ref{ncco} shows $\Delta \lambda ( T)$ measured in a single crystal of
NCCO. The inset shows the low-temperature range. Below $T \approx 4$ K there
is a pronounced upturn, which we attribute to the paramagnetic contribution of
Nd$^{3+}$ ions. The upper solid line in the inset to Fig.\ref{ncco} shows the
power law fit ($\lambda_{L} \propto T^{n}$) which yields $n=1.35 \pm 0.03$ and
$n=1.40 \pm 0.03$ for $C=0.3$ and $0.05$, respectively. (Fits for the two
different values of $C$ are indistinguishable on this scale.) The lower line
shows analogous fits to the low temperature $s-$wave expression from which we
obtain $\Delta (0)/T_c=0.569 \pm 0.006$ and $\Delta (0)/T_c=0.573 \pm 0.006$.
Fits were obtained from data up to $t=0.32$. The higher temperature data is
shown for completeness. Changing the value of $\Theta$ from 1.2 to 2 also had
a small effect on the fit parameters. The value of $\Delta (0)/T_c$ is close
to that obtained in PCCO. Again, for a strict test we should fit over a
correspondingly reduced temperature range. However, the dominant paramagnetic
contribution below $t < 0.16$ renders this procedure meaningless. The n = 1.4
exponent obtained from the power law fit is closer to the clean d-wave limit
and could imply that after correction for paramagnetism, unitary-limit
scattering in NCCO is smaller than in PCCO.

In conclusion, we have measured the penetration depth $\lambda ( T )$ in
electron-doped PCCO and NCCO single crystals down to 0.4 K. In non-magnetic
PCCO, $\rho_s$ decreases quadratically with temperature up to $t \approx $ 0.3,
consistent with a dirty d-wave (gapless) scenario. The correlation between
$T_c$ and the rate of change of superfluid density is also consistent with
this picture. In NCCO, a large paramagnetic contribution to the penetration
depth was observed. $\lambda_L ( T )$ was found to vary as $T^{1.4}$. For both
materials, a fit over the same temperature range to an s-wave model sets an
upper limit of $\Delta (0)/T_c = 0.57$ but requires unrealistically small
values $\lambda (0)$. For PCCO, the test can be made more stringent and we
reduce the upper limit to $\Delta_{min} (0)/T_c = 0.2$.

We thank J. R. Cooper, A. A. Abrikosov, S. M. Anlage, N. Goldenfeld, P. J.
Hirschfeld, R. Klemm, L. Taillefer, A. Mourachkine, Y. Yeshurun and C. C.
Tsuei for useful discussions and communications. This work was supported by
the Science and Technology Center for Superconductivity Grant No. NSF-DMR
91-20000. Work in Maryland is supported by Grant No. NSF-DMR 97-32736.

\end{multicols}

\end{document}